\documentclass[prl,aps,amssymb,amsmath]{article}
\usepackage{amssymb,amsmath}

\begin{document}
\date{July 29, 2003}
\title{{\bf Generally covariant Quantum Mechanics\footnote{Presented at
the {\bf Symmetry 2003} conference in Kiev.}}}
\author{Jaros{\l}aw Wawrzycki\footnote{Electronic address: 
Jaroslaw.Wawrzycki@ifj.edu.pl or  
jwaw@th.if.uj.edu.pl}
\\Institute of Nuclear Physics, ul. Radzikowskiego 152, 
\\31-342 Krak\'ow, Poland}
\maketitle
\newcommand{\ud}{\mathrm{d}}
\begin{abstract}
We present a complete theory, which is a generalization of Bargmann's theory
of factors for ray representations. We apply the theory to the generally
covariant formulation of the Quantum Mechanics.
\end{abstract}

\section{Problem}

In the standard Quantum Mechanics (QM) and the Quantum Field Theory (QFT) 
the spacetime coordinates are pretty classical variables. Therefore
the question about the general covariance of QM and QFT
emerges naturally just like in the classical theory:
\begin{enumerate} 

\item[]
        {\bf what is the effect of a changing of the spacetime coordinates
        in QM and QFT when the changing does not form any symmetry
        transformation?}

\end{enumerate}

It is a commonly accepted believe that there are no substantial 
difficulties if we refer the question to the wave equation. We simply
treat the wave equation, and do not say why, in such a manner as if it 
was a classical equation. The only problem arising is to find the 
transformation rule $\psi \to T_{r}\psi$ for the wave function $\psi$.
This procedure, which on the other hand can be seriously objected, does
not solve the above stated problem. The heart of the problem as well
as of QM and QFT lies in the Hilbert space of states and just in 
finding the representation $T_{r}$ of the covariance group in question.
The trouble gets its source in the fact that the covariance transformation
changes the form of the wave equation such that $\psi$ and $T_{r}\psi$
do not belong to the same Hilbert space, which means that $T_{r}$ does not
act in the ordinary Hilbert space. This is not compatible with the 
paradigm worked out in dealing with symmetry groups.

\section{Resume of the reformulation of the Quantum Mechanics}

We have shown that covariance group acts in a Hilbert bundle
$\mathcal{R}\triangle \mathcal{H}$ over the time in the nonrelativistic 
theory and in a Hilbert bundle $\mathcal{M}\triangle \mathcal{H}$ over
the spacetime $\mathcal{M}$ in the relativistic case. The wave functions 
are the appropriate cross sections of the bundle in question. The exponent
$\xi(r,s,p)$ in the formula
\begin{displaymath}
T_{r}T_{s} = e^{i\xi(r,s,p)}T_{rs},
\end{displaymath}  
depends on the point $p$ of the base of the bundle in question: 
that is, $\xi$ depends on the time $t$ in the nonrelativistic 
theory and on spacetime point $p$ in the relativistic theory if 
there exists a nontrivial gauge freedom. 

Moreover, we argue that the bundle $\mathcal{M}\triangle \mathcal{H}$
is more appropriate for treating the covariance as well as the symmetry
groups then the Hilbert space itself. Namely, we show that from the 
more general assumption that the representation $T_{r}$ of the Galilean
group acts in $\mathcal{R}\triangle \mathcal{H}$ and has an exponent
$\xi(r,s,t)$ depending on the time $t$ we reconstruct the nonrelativistic
Quantum Mechanics. Even more, in the less trivial case of the theory
with nontrivial time-dependent gauge describing the spin less quantum 
particle in the Newtonian gravity we are able to infer the wave equation
and prove the equality of the inertial and gravitational masses.  
  
In doing it we apply extensively the classification theory for exponents
$\xi(r,s,t)$ of $T_{r}$ acting in $\mathcal{R}\triangle \mathcal{H}$ 
and depending on the time. 

In the presented theory which is slightly more general then the standard one
the gauge freedom emerges from the very nature of the fundamental laws of
Quantum Mechanics. By this it opens a new perspective in solving the
troubles in QFT caused by the gauge freedom. 

\vspace{1ex}

{\bf Interpretation.} \,\, The physical interpretation ascribed to the cross 
section $\psi$ is as follows. Each experiment is, out of its very nature, a 
spatiotemporal event. To each act of measurement carried out at the 
spacetime point $p_{0}$ we ascribe a self-adjoint operator $Q_{p_{0}}$ acting 
in the Hilbert space $\mathcal{H}_{p_{0}}$ and ascribe to the spectral 
theorem for $Q_{p_{0}}$ the standard interpretation. Hence, assuming for 
simplicity that $Q_{p_{0}}$ is bounded, if $\phi_{0} \in \mathcal{H}_{p_{0}}$ 
and $\lambda_{0} =\lambda_{o}(p_{0})$ is a characteristic vector and its 
corresponding characteristic value of $Q_{p_{0}}$  respectively then we have 
the following statement.  
\emph{If the experiment corresponding to} $Q_{p}$  
\emph{was performed at the spatiotemporal event} $p_{0}$
\emph{on a system in the state described
by the cross section $\psi$, then the probability of the measurement
value to be $\lambda_{0}(p_{0})$ and the system to be found
in the state described by $\phi$ such that $\phi(p_{o})=\phi_{0}$
after the experiment is given by the square of the absolute value of 
the Borel function $\vert f_{\psi}(p_{0},\phi_{0}) \vert^{2} 
= \vert (\phi_{0},\psi_{p_{0}})\vert^{2}$  
induced by the cross section $\psi$}. In the nonrelativistic 
case the above statement is a mere rephrasing of the well
established knowledge. 

\vspace{1ex}

See \cite{JW} for detail treatment, where the theory was proposed.

\section{Resume of the generalization of the Bargmann's theory 
of factors}

Our main task was to construct the general classification
theory of spacetime dependent exponents $\xi(r,s,p)$ of representations
acting in $\mathcal{M}\triangle \mathcal{H}$, see \cite{JW}. On the other 
hand the presented theory can be viewed as a generalization of the 
Bargmann's \cite{Bar} classification theory of exponents $\xi(r,s)$
of representations acting in ordinary Hilbert spaces,
which are independent of $p \in \mathcal{M}$.

\vspace{1ex}

{\bf Definition.} \,\, By an \emph{isomorphism} of the Hilbert bundle 
$\mathcal{M}\triangle\mathcal{H}$ with the Hilbert bundle
$\mathcal{M}'\triangle\mathcal{H}'$ we shall mean a Borel isomorphism $T$
of $\mathcal{M}\triangle\mathcal{H}$ on $\mathcal{M}'\triangle\mathcal{H}'$ 
such that for each $p \in \mathcal{M}$ the restriction of $T$ to
$p \times \mathcal{H}_{p}$ has some $q \times \mathcal{H}_{q}'$
for its range and is unitary when regarded as a map of $\mathcal{H}_{p}$
on $\mathcal{H}_{q}'$. The induced map carrying $p$ into $q$ is clearly
a Borel isomorphism of $\mathcal{M}$ with $\mathcal{M}'$ and we denote it
by $T^{\pi}$. The above defined $T$ is said to be an \emph{automorphism}
if $\mathcal{M}\triangle\mathcal{H}=\mathcal{M}'\triangle\mathcal{H}'$. 
Note that for any automorphism $T$ we have 
$(T\psi,T\phi)_{T^{\pi}p}=(\psi,\phi)_{p}$, but in general
$(T\psi,T\phi)_{p}\neq(\psi,\phi)_{p}$. By this any automorphism $T$ is
what is frequently called a \emph{bundle isometry}. (We use the Hilbert 
bundle with the ordinary Borel structure in the total space and with 
the ordinary manifold structure in the base $\mathcal{M}$, 
see e.g. \cite{Mackey}.) 

{\bf Definition.} \,\, The function $r \to T_{r}$ from a group $G$ into the 
set of automorphisms (bundle isometries) of $\mathcal{M}\triangle\mathcal{H}$ 
is said to be a \emph{general factor representation} of $G$ associated 
to the action $G \times \mathcal{M} \ni r,p \to r^{-1}p \in \mathcal{M}$ of
$G$ on $\mathcal{M}$ if $T_{r}^{\pi}(p) \equiv r^{-1}p$ for all $r \in G$,
and $T_{r}$ satisfy the condition 
\begin{displaymath}
T_{r}T_{s} = e^{i\xi(r,s,p)}T_{rs}.
\end{displaymath}

Here we give only the summing up

{\bf Theorem.}\,\, (1) On a Lie group $G$, every local exponent $\xi(r,s,p)$ 
is equivalent to a canonical
 local exponent $\xi'(r,s,p)$  which, 
on some canonical neighborhood ${\mathfrak{N}}_{0}$, is analytic
in canonical coordinates 
of $r$ and $s$ and and vanishes if $r$ 
and $s$ belong to the same one-parameter subgroup. Two canonical
local exponents $\xi,\xi'$ are equivalent if and only if 
$\xi'(r,s,p) = \xi(r,s,p) + \Lambda(r,p)+\Lambda(s,r^{-1}p)-\Lambda(rs,p)$ 
on some canonical
neighborhood, where $\Lambda(r,p)$ is a linear form in the canonical 
coordinates of $r$ such that
 $\Lambda(r,sp)$ does not depend on $s$
if $s$ belongs to the same one-parameter subgroup as $r$.
(2) To every canonical
 local exponent of $G$ corresponds uniquely an 
infinitesimal exponent $\Xi(a,b,p)$ on the Lie
 algebra $\mathfrak{G}$ of 
$G$, i.e. a bilinear antisymmetric form which satisfies  the 
identity\footnote{$\boldsymbol{a}\Xi(b,c,p)$ stands for the diffrential 
operator $d/d\tau \Xi(b,c,(\tau a)p)\vert_{\tau=0}$.}
$\Xi([a,a'],a'',p) +\Xi([a',a''],a,p)+ \Xi([a'',a],a',p) 
= \boldsymbol{a}\Xi(a',a'',p) + \boldsymbol{a'}\Xi(a'',a,p) +
\boldsymbol{a''}\Xi(a,a',p)$. The correspondence is linear. (3) Two canonical 
local exponents 
$\xi,\xi'$ are equivalent if and only if the corresponding 
$\Xi$, $\Xi'$ are equivalent, i.e.\footnote{$\tau\to\tau a$ is a one-parameter
group generated by $a \in \mathfrak{G}$.} 
$\Xi'(a,b,p) = \Xi(a,b,p) 
+ \boldsymbol{a}\Lambda(b,p) - \boldsymbol{b}\Lambda(a,p) - \Lambda([a,b],p)$
where $\Lambda(a,p)$ is a linear form in $a$ on $\mathfrak{G}$ such that 
$\tau \to \Lambda(a,(\tau b)p)$ is constant if $a = b$.
(4) There exist a one-to-one correspondence between the equivalence classes 
of local exponents
 $\xi$ (global in $p \in \mathcal{M}$) of $G$ and the 
equivalence classes of infinitesimal exponents $\Xi$ of 
$\mathfrak{G}$.

\section{The physical motivation}\label{motivation}

It will be instructive to investigate 
the problem for the free particle in the flat Galilean spacetime. The set of 
solutions $\psi$ of the Schr\"odinger equation which are admissible in Quantum 
Mechanics is 
precisely given by 
\begin{displaymath}
\psi(\vec x,t)=(2\pi)^{-3/2} \int \varphi(\vec k)e^{-i\frac{t}{2m}\vec{k}
\centerdot \vec k+i\vec k\centerdot\vec x}
\, {\ud}^{3}{k},
\end{displaymath}
where $p=\hslash k$ is the linear momentum and $\varphi(\vec{k})$ is any square 
integrable function. The functions $\varphi$ (wave functions in the "Heisenberg 
picture") form a Hilbert space $\mathcal{H}$ with the inner product
\begin{displaymath}
(\varphi_{1}, \varphi_{2})=\int \varphi_{1}^{*}(\vec{k})\varphi_{2}(\vec{k}) \, 
{\ud}^{3}{k}.
\end{displaymath}
The correspondence between $\psi$ and $\varphi$ is one-to-one. 

But in general the construction fails if the Schr\"odinger equation possesses 
a nontrivial gauge freedom.
We explain it. 

We need not to use the 
Fourier transform. {\bf What is the role 
of the Schr\"odinger equation in the above construction of $\mathcal{H}$?}
In the above construction the Hilbert space 
$\mathcal{H}$ is isomorphic to the 
space of square integrable functions 
$\varphi(\vec{x})\equiv \psi(\vec{x},0)$ -- the set of square integrable
space of initial data for the Schr\"odinger equation. 
The connection between $\psi$ and
$\varphi$ is given by the time evolution 
$U(0,t)$ operator (by the Schr\"odinger equation):
\begin{displaymath}
U(0,t)\varphi=\psi.
\end{displaymath}
The correspondence between $\varphi$ and $\psi$ has all formal properties such 
as in the above Fourier 
construction. 
 Denote the space of the initial square integrable 
data $\varphi$ on the simultaneity
hyperplane $t(X)=t$ by ${\mathcal{H}}_{t}$. 
The space of wave functions $\psi(\vec{x},t) = U(0,t)\varphi(\vec{x})$
isomorphic to the Hilbert space ${\mathcal{H}}_{0}$ of $\varphi$'s is called in 
the common "jargon" the 
"Schr\"odinger picture".  

However, the connection between $\varphi(\vec{x})$ and $\psi(\vec{x},t)$ is not 
unique in general,
if the wave equation possesses a gauge freedom. Namely, 
consider 
the two states $\varphi_{1}$ and $\varphi_{2}$ and ask the question: 
when the two states are equivalent and by this indistinguishable? The answer is 
as follows: they are equivalent if 
\begin{displaymath}
\vert(\varphi_{1},\varphi)\vert \equiv \Big\vert\int \psi_{1}^{*}(\vec{x},t)
\psi(\vec{x},t) \, {\ud}^{3}{x}\Big\vert =
\vert(\varphi_{2},\varphi)\vert \equiv
\end{displaymath}

\begin{equation}\label{row}
\equiv \Big\vert\int \psi_{2}^{*}(\vec{x},t)\psi(\vec{x},t) \, 
{\ud}^{3}{x}\Big\vert,
\end{equation}
for any state $\varphi$ from $\mathcal{H}$, or for all $\psi=U\varphi$ 
($\psi_{i}$ are defined to be = 
$U(0,t)\varphi_{i}$). Substituting $\varphi_{1}$ 
and then 
$\varphi_{2}$ for $\varphi$ and making use of the Schwarz's inequality 
one gets: $\varphi_{2}=e^{i\alpha}\varphi_{1}$,
where $\alpha$ is any 
constant\footnote{This gives the conception of the ray, introduced to Quantum 
Mechanics by Hermann Weyl \cite{Weyl}: a physical 
state does not correspond uniquely to a normed state $\varphi \in \mathcal{H}$, 
but it is uniquely described by a 
\emph{ray}, two states belong to the same 
ray if they differ by a constant phase factor.}.
The situation for $\psi_{1}$ and $\psi_{2}$ is however different. 
In general the condition (\ref{row}) is fulfilled if 
\begin{displaymath}
\boldsymbol{\psi_{2}=e^{i\Lambda(t)}\psi_{1}}
\end{displaymath}
{\bf and the phase factor can depend on time}. Of course it has to be consistent 
with the wave equation, that is, together
with a solution $\psi$ to the wave 
equation the wave function $e^{i\Lambda(t)}\psi$ also is a solution to the appropriately
gauged wave equation. \emph{A priori}
one can not exclude the existence of such 
a consistent time evolution. This is not a new observation, it was 
noticed by John von Neumann\footnote{J. v. Neumann, {\sl Mathematical Principles 
of Quantum Mechanics}, University Press, Princeton (1955). He did not mention 
about the gauge freedom on that occasion. But the gauge freedom 
is necessary for the equivalence of $\psi_{1}$ and $\psi_{2} 
= e^{i\Lambda(t)}\psi_{1}$.}, but it seems that it has never 
been deeply investigated
(probably because the ordinary nonrelativistic 
Schr\"odinger equation has a gauge symmetry 
with constant $\Lambda$). 
The space of waves $\psi$ describing the system cannot be reduced in the above 
way to any fixed Hilbert space $\mathcal{H}_{t}$ with a fixed $t$. So, 
the existence of the nontrivial gauge freedom leads to the 

\vspace{1ex}

{\bf Hypothesis}. \emph{The two waves $\psi$ and $e^{i\Lambda(t)}\psi$ are
quantum-mechanically indistinguishable}.

\vspace{1ex}

Moreover, we are obliged to use the whole Hilbert bundle $\mathcal{R}\triangle 
\mathcal{H}: t \to \mathcal{H}_{t}$ over the time instead of a fixed Hilbert
space $\mathcal{H}_{t}$, with the appropriate cross sections as the waves
$\psi$. 

Consider now an action $T_{r}$ of a group $G$ in the space of waves $\psi$.    
From our analysis it follows that it is natural to replace the ordinary
postulate:

\vspace{1ex}

{\bf Classical-like  postulate}. \emph{The group $G$ is a symmetry 
group if and only if the wave equation is invariant under the transformation
$x' = rx, r\in G$ of independent variables and the transformation $\psi'= 
T_{r}\psi$ of the wave function}.

\vspace{1ex}
 
by the more appropriate alternative:

\vspace{1ex}

{\bf Quantum postulate}. \emph{The group $G$ is a symmetry 
group if and only if  the transformation $x' = rx, r\in G$ of independent 
variables and the transformation $\psi'= T_{r}\psi$ of the wave function 
transform the wave equation into a gauge-equivalent one}.

\vspace{1ex}

Acceptation of the {\bf Quantum postulate} gives a new perspective 
for solving the two very difficult problems \cite{JW}:
\begin{enumerate}

\item[{\bf (a)}]
          {\bf generally covariant formulation of Quantum Mechanics},

\item[{\bf (b)}]
          {\bf the troubles in the Quantum Field Theory caused by the 
          gauge freedom}.

\end{enumerate}
Moreover, with the help of the {\bf Hypothesis} we can see that both
{\bf (a)} and {\bf (b)} are deeply connected \cite{JW}.

\section{Connection to the troubles with the gauge freedom}\label{gauge}

Now, we return to the problem {\bf (b)}.
It should be mentioned at this place that the troubles in QFT 
generated
by the gauge freedom are of general character, and are well known. 
For example, there do not exist vector particles with 
helicity = 1, which is a consequence
of the theory of unitary 
representations of the Poincar\'e group, as was shown by J. {\L}opusza\'nski 
\cite{Lopuszanski}. This is apparently in contradiction with the existence of 
vector particles with
helicity = 1 in nature -- the photon, which is connected 
with the electromagnetic four-vector potential.
The connection of the problem 
with the gauge freedom is 
well known \cite{Lopuszanski}. We omit however the 
difficulty if we allow the inner product in
the "Hilbert space" to be not 
positively defined, see  \cite{Gupta}, 
or  \cite{BRST}.
Due to \cite{Lopuszanski}, the vector potential (promoted to be an 
operator valued distribution in QED)
cannot be a vector field, if one 
wants to have the inner product positively defined then together with
the coordinate transformation the gauge transformation has to be applied, 
which breaks the vector
character of the potential. Practically it means that 
any gauge condition which brings the theory into
the canonical form such that 
the quantization procedure can be consequently applied (with the positively
defined inner product in the Hilbert space) breaks the four-vector character 
of the electromagnetic potential, the 
Coulomb gauge condition is an example. 
To achieve the Poincar\'e symmetry of Maxwell
equations with such a gauge 
condition (the Coulomb gauge condition for example), it is impossible to preserve 
the vector character 
of the potential -- together with the coordinate 
transformation a well defined (by the coordinate transformation)
gauge 
transformation $f$ has to be applied:
\begin{displaymath}
A_{\mu} \to A'_{\mu'} = \frac{\partial x^{\nu}}{\partial x^{\mu'}}(A_{\nu} 
+ \partial_{\nu}f).
\end{displaymath}
This means that the electromagnetic potential can form
a generalized ray representation $T_{r}$ of the Poincar\'e 
group at most, with
 the spacetime-dependent factor $e^{i\xi}$ if the scalar 
product is positively defined. 
One may ask: how possible is it if the 
Poincar\'e
 group is not only a covariance group but at the same time a 
symmetry group? The solution of this paradox
 on the grounds of the 
existing theory is rather obscure. We propose the following solution.
The factor $e^{i\xi}$ is spacetime independent for the symmetry group 
but under the assumption 
that the fundamental space describing the states of a quantum system is 
 the ordinary Hilbert space and the {\bf Classical-like postulate} is true.
But we have presented serious objections to this assumption. 
Moreover, the nonrelativistic quantum theory can be 
reconstructed from the more general assumption about the space of quantum 
mechanical states saying that it compose the space of appropriate cross sections
of the Hilbert bundle $\mathcal{R}\triangle\mathcal{H}$ 
over time $t \in \mathcal{R}$. The Schr\"odinger equation can be uniquely 
reconstructed from the generalized ray representations of the Galilean group. 
We watch for also a more fundamental justification of this assumption in the 
presumption that the time is a purely classical variable in the nonrelativistic 
quantum mechanics or so to speak a parameter. The most general unitary
representation of the locally compact commutative group of the time 
real line acts in a Hilbert bundle $\mathcal{R}\triangle\mathcal{H}$  
over the time, see Mackey \cite{Mackey}. So, the assumption about the 
"classicity" of the time $t$ fixes the structure of space of quantum states 
to be a subset of cross sections of a Hilbert bundle over the time.  
This is the peculiar property of the Galilean group 
structure that the whole construction degenerates as if we were started from
the ordinary ray representation over the ordinary Hilbert space and
the theorem that the pase $e^{i\xi}$ is spacetime independent is true in 
this case, but only accidentally.  
The generalization to the relativistic case is natural. First we postulate
the spacetime coordinates to be classical commutative variables, 
which leads to the Hilbert bundle $\mathcal{M}\triangle\mathcal{H}$  
over the space-time manifold $\mathcal{M}$. The factor of the representation 
of the Poincar\'e group acting in the bundle $\mathcal{M}\triangle\mathcal{H}$ 
has not to be a constant with respect to space-time coordinates even when it 
is a symmetry group. This solves the paradox.

On the other hand H. Weyl (pages 272-276  of the Dover ed. of \cite{Weyl})  
showed that the Quantum Kinematics represented by the Heisenberg commutation
relations is nothing else but a unitary factor representation of an Abelian
group. This Weyl' construction can be viewed as emerging from the gauge 
freedom with the constant phase (constant $\Lambda$), i.e. a special case
of the Bargmann' theory, see \cite{Bar}. This is just the case of the Heisenberg'
analysis of commutation relations for the atom which is determined (in the first  
approximation) by the Coulomb field of its nuclei charge. That is, the field 
which is naturally connected with the constant phase gauge freedom. But the 
general electromagnetic interactions possess the spacetime dependent gauge 
freedom. So, by the construction our generalization seems to be natural
for treating the quantum electromagnetic interactions.

\section{Connection to the troubles with the position operator}

It was shown in section {\bf \ref{motivation}} that our generalization
with states as cross sections in a Hilbert bundle 
$\mathcal{M}\triangle \mathcal{H}$ over the whole spacetime $\mathcal{M}$  
looses physical sense in the nonrelativistic theory. In this case we have
the bundle $\mathcal{R}\triangle \mathcal{H}$ over the time. It should
be noted that we can recover this fact from the more general point of view.
Namely, any exponent $\xi(r,s,p)$ of a generalized factor representation of 
the Galilean group acting in $\mathcal{M}\triangle \mathcal{H}$ is equivalent
to a time dependent one: $\xi(r,s,p) = \xi(r,s,t)$, such that we can restrict
the whole analysis to the bundle  $\mathcal{R}\triangle \mathcal{H}$ over 
the time. This is a peculiar property of the Galilean group. This is not the
case for the Poincar\'e group of course. The time, being a classical variable,
does not posses any quantum mechanical measurement operator. But, contrary
to the time, the spacetime coordinates do possess operators in the nonrelativistic 
theory. In the relativistic case all the spacetime coordinates play the role
of the time and do not possess quantum mechanical operators. This is in agreement
to the standard problem in construction of the position operator in the 
relativistic theory, compare e.g. \cite{Mackey} and \cite{Wigner}.


\begin{thebibliography}{99}
\footnotesize

\vspace{.5cm}

\bibitem{JW} J. Wawrzycki, math-ph/0301005.
\bibitem{Bar} V. Bargmann, Ann. Math. {\bf 59}, 1, (1954). 
\bibitem{Mackey} G. W. Mackey, {\sl Unitary Group Representations in 
Physics, Probability, and Number Theory}. Addison-Wesley Publishing Company,
INC. The Advanced Book Program. Redwood City-California, New York, Amsterdam,
Wokingham-UK (1989).  
\bibitem{Weyl} H. Weyl, {\sl Gruppentheorie und Quantenmechanik}, 
Verlag von S. Hirzel in Leipzig (1928).
\bibitem{Lopuszanski} J. {\L}opusza\'nski, Fortschritte der Physik {\bf 26}, 
261, (1978);
{\sl Rachunek spinor\'ow}, PWN, Warszawa 1985 (in Polish).
\bibitem{Gupta} S. N. Gupta, Proc. Phys. Soc. {\bf 63}, 681, (1950); 
K. Bleuler, Helv. Phys. Acta {\bf 23}, 567, (1950).
\bibitem{BRST} C. Becchi, A. Rouet and R. Stora, Commun. Math. Phys. 
{\bf 42}, 127, (1975);
Ann. Phys. {\bf 98}, 287, (1976); I. V. Tyutin, 
Liebiediev Institute preprint N39(1975). See for example:
S. Weinberg, {\sl The Quantum Theory of Fields}, volume II, Univ. Press, 
Cambridge 1996, where 
the BRST-formalism is described.
\bibitem{Wigner} E. P. Wigner, {\sl Interpretation of Quantum Mechanics}, 
Princeton University Press, Princeton (1983).
\end{thebibliography}
\end{document}